\documentclass[prl,twocolumn,showpacs,floatfix,amssymb,amsmath,superscriptaddress]{revtex4}

\usepackage{graphicx}

\begin{document}

\title[Molecular Synchronization Waves]{Molecular Synchronization Waves in
Arrays of Allosterically Regulated Enzymes}
\author{Vanessa Casagrande}
\affiliation{Hahn-Meitner-Institut, Glienicker Stra{\ss}e 100, 14109 Berlin, Germany}
\author{Yuichi Togashi}
\altaffiliation[Present address: ]{Nanobiology Laboratories, Graduate School of Frontier Biosciences, Osaka University, 1-3 Yamadaoka, Suita, Osaka 565-0871, Japan}
\email{togashi@phys1.med.osaka-u.ac.jp}
\affiliation{Fritz-Haber-Institut der Max-Planck-Gesellschaft, Faradayweg 4-6, 14195 Berlin, Germany}
\author{Alexander S. Mikhailov}
\email{mikhailov@fhi-berlin.mpg.de}
\affiliation{Fritz-Haber-Institut der Max-Planck-Gesellschaft, Faradayweg 4-6, 14195 Berlin, Germany}

\begin{abstract}
Spatiotemporal pattern formation in a product-activated enzymic
reaction at high enzyme concentrations is investigated. Stochastic simulations
show that catalytic turnover cycles of individual enzymes can become coherent
and that complex wave patterns of molecular synchronization can develop.
The analysis based on the mean-field approximation indicates that the observed
patterns result from the presence of Hopf and wave bifurcations in the
considered system.
\end{abstract}

\pacs{82.40.Ck, 87.18.Pj, 82.39.Fk, 05.45.Xt}

\maketitle

Molecular machines, such as molecular motors, ion pumps and some enzymes,
play a fundamental role in biological cells and can be also used in the
emerging soft-matter nanotechnology \cite{Kinbara}. A protein machine is a
cyclic device, where each cycle consists of conformational motions initiated
by binding of an energy-bringing ligand \cite{Blumenfeld,Gerstein}.
In motors, such internal motions generate mechanical work \cite{Prost-review},
while in enzymes they enable or facilitate chemical reaction events (see,
e.g., \cite{Lerch1,Lerch2}). Much attention has been attracted to
studies of biomembranes with ion pumps and molecular motors,
where membrane instabilities
and synchronization effects have been analyzed \cite{Prost1,Prost2,Chen}.
Here, a different class
of distributed active molecular systems --- formed by enzymes --- is considered.
The catalytic activity of an allosteric enzyme protein is activated or
inhibited by binding of small regulatory molecules; the role
of such regulatory molecules can be played by products of the same
reaction \cite{Goldbeter}. Previous investigations of simple
product-regulated enzymic systems \cite{Stange1998,Stange1999} and
enzymic networks \cite{Ouyang} in small spatial volume with full diffusional
mixing have shown that spontaneous synchronization of molecular turnover
cycles can take place there. External molecular synchronization of enzymes of
the photosensitive P-450 dependent monooxygenase system by periodic optical
forcing has been experimentally demonstrated \cite{Gruler}.

In this Letter, spatiotemporal pattern formation in enzymic arrays is
investigated. In such systems, immobile enzymes are attached to a
solid planar support immersed into a solution through which fresh substrate
is supplied and product molecules are continuously removed. Product
molecules released by an enzyme diffuse through the solution and activate
catalytic turnover cycles of neighbouring enzymes in the array.

\begin{figure}
\includegraphics[width=40mm]{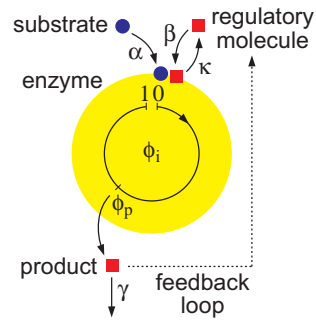}
\caption{(Color online) A sketch of the model.}
\label{Enzyme}
\end{figure}

A simple stochastic model \cite{Stange1999} of an enzyme as a cyclic machine
(a stochastic phase oscillator), shown in Fig. 1, is used. Binding of a
substrate molecule to an enzyme $i$ initiates an ordered internal
conformational motion, described by the
conformational phase coordinate $\phi_{i}$. The initial state corresponds
to the phase $\phi_{i}=0$. The catalytic conversion event takes place
and the product is released at the state $\phi_{p}$ inside
the cycle. After that, the conformational motion continues until the
equilibrium state of the enzyme ($\phi_{i}=1$) is finally reached.
Initiation of a turnover cycle is a random event, occurring at a certain
probability rate.
We assume that substrate is present in abundance,
and its concentration is not affected by the reactions.
Conformational motion inside the cycle is modeled as a stochastic
diffusional drift process, described by equation
$\overset{.}{\phi}_{i}=v+\eta_{i}(t)$,
where $v$ is the mean drift velocity and $\eta_{i}(t)$ is an
internal white noise with
$\left\langle \eta_{i}(t)\eta_{j}(t^{\prime})\right\rangle = 2\sigma \delta_{ij}\delta(t-t^{\prime})$ where
$\sigma$ specifies intensity of
intramolecular fluctuations.

Allosterically activated enzymes possess a site on their
surface where regulatory molecules can become bound. Binding of a
regulatory molecule leads to conformational change that enhances
catalytic activity of the enzyme. A regulatory molecule binds to
an enzyme with rate constant $\beta$ and dissociate from it with
rate constant $\kappa$. Binding of a regulatory molecule at an enzyme raises
its probability to start a cycle from $\alpha_{0}$ to $\alpha_{1}$. We
assume that a regulatory molecule can bind to an enzyme only in its rest
state and this molecule is released when the cycle is started. The role of
regulatory molecules is played by product
molecules of the same reaction. Immobile enzymes are randomly distributed in
space with concentration $c$.
Product diffuses at diffusion constant $D$ and undergoes decay at
rate constant $\gamma$. The characteristic diffusion length of product
molecules is $l_{diff} = \sqrt{D/\gamma}$.

In our stochastic 2D simulations, the medium was discretized into spatial cells
(up to $256 \times 256$), each containing a number of
enzyme molecules. The cells were so small that diffusional mixing of product
molecules in a cell within the shortest characteristic time of the reaction
could always take place. Each enzyme was described by the stochastic model
given above; diffusion of product molecules was modeled as a random walk
over a discrete cell lattice. The mean cycle time $\tau =1/v$ was chosen
as the time unit ($\tau = 1$).
Systems including up to 655 360
enzymes were used in the simulations.

\begin{figure}
\includegraphics[width=86mm]{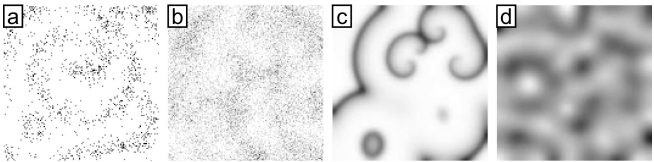}
\caption{Stochastic (a,b) and mean-field (c,d) simulations of 2D wave patterns;
(a) $\tau_{p}=0.14$, $c=1$, and $\beta=300$,
(b) $\tau_{p}=0.25$, $c=10$, and $\beta=10$,
(c) $\tau_{p}=0.14$, $c=1$, and $\beta=300$,
(d) $\tau_{p}=0.34$, $c=100$, and $\beta=1.42$.
Other parameters are $\alpha_{0}=1$, $\alpha_{1}=1000$,
$\kappa=10$, $\gamma=10$, $\sigma=0$, $D=100$.
The linear size of the shown area is $L = 40$ $l_{diff}$ in all panels.}
\label{patterns-2d}
\end{figure}

Figure 2a,b (see also Videos 1 and 2
in ref. \cite{epaps})
shows two typical examples of stochastic 2D simulations. Here, spatial
distributions of product molecules are displayed. Waves of product
concentration are propagating through the medium. In a peak of a wave, many
locally present enzymes are simultaneously releasing product molecules.
Since product release can take place only at a certain stage inside the
cycle, this means that the cycles of enzymes are locally synchronized. Not
only regular wave structures, such as rotating spiral waves or target
patterns (Fig. 2a), but also complex regimes of wave turbulence (Fig. 2b)
have been observed.

To understand and interpret stochastic simulation results, an analytical
study of the system in the mean-field approximation, which holds in the
limit of high enzyme concentrations, has been
performed. In this approximation, the system is characterised by three
continuous variables $n_{0}(\mathbf{r},t)$, $n_{1}(\mathbf{r},t)$ and
$m(\mathbf{r},t)$ which represent local concentrations of enzymes in the rest
state without or with regulatory molecules attached ($n_{0}$ and $n_{1}$)
and local concentration of the product ($m$). For simplicity, internal
fluctuations in enzymes are neglected ($\sigma = 0$). Thus, all enzymes which
have started their cycles at some time $t$ would release their products at a
definite time $t+\tau_{p}$ (with $\tau_{p}=\phi_{p}/v$) and finish their
cycles, returning to the rest state, at time $t+\tau$. Therefore, the
system is described by a set of three reaction-diffusion equations with time
delays,
\begin{subequations}
\begin{eqnarray}
\frac{\partial n_{1}}{\partial t} &=& \beta mn_{0}-\kappa n_{1}-\alpha_{1}n_{1} \\
\frac{\partial n_{0}}{\partial t} &=& -\beta mn_{0}+\kappa n_{1}-\alpha_{0}n_{0} + \alpha_{0}n_{0}(t-\tau) \notag \\
&& + \alpha_{1}n_{1}(t-\tau) \\
\frac{\partial m}{\partial t} &=& -\beta mn_{0}+\kappa n_{1}+\alpha_{1}n_{1}-\gamma m + \alpha_{0}n_{0}(t-\tau_{p}) \notag \\
&& + \alpha_{1}n_{1}(t-\tau_{p}) + D\nabla^{2}m.
\end{eqnarray}

The system always has a uniform stationary state with certain
concentrations $\overline{n}_{0}$, $\overline{n}_{1}$ and $\overline{m}$, which
can be found as solutions of the respective algebraic equations. This state
corresponds to the absence of synchronization. However, it may become
unstable if allosteric activation is strong enough. To analyze
stability, small perturbations $\delta n_{0}$, $\delta n_{1}$ and $\delta m$
are added to the stationary state, equations (1) are linearized and
their solutions are sought as
$\delta n_{0} \sim \delta n_{1} \sim \delta m \sim \exp \left( \lambda_{q}t-iqx \right)$ with
$\lambda_{q}=\mu_{q}+i\omega_{q}$. Thus, each spatial mode with
wavevector $q$ is characterized by its frequency $\omega_{q}$ and its rate
of growth $\mu_{q}$. The properties $\mu_{q}$ and $\omega_{q}$ are given by
the roots of a characteristic equation which is determined by the
linearization matrix of equations (1). The steady state becomes unstable
when at least one spatial mode with some wavenumber $q_{0}$ starts to grow
($\mu_{q_{0}}>0$).

As the bifurcation parameter, coefficient $\beta$ can be chosen. If
regulatory molecules cannot bind to enzymes ($\beta = 0$), feedback is absent
and instabilities are not possible. On the other hand, allosteric activation
becomes strong if regulatory molecules can easily bind and, in this case,
emergence of oscillations and wave patterns can be expected.
Our bifurcation analysis reveals that,
depending on the parameters of the system,
it can exhibit either a Hopf or a wave bifurcation \cite{Thesis}.
As a result of the Hopf bifurcation, uniform
oscillations with $q=0$ develop. Because of the presence of delays in
equations (1), the characteristic equation is nonpolynomial in terms of
$\lambda$ and, generally, a number of oscillatory solutions with different
frequencies $\omega$ are possible. Physically, such solutions correspond to
formation of several synchronous enzymic groups. This effect has been
previously extensively investigated for similar systems in small spatial
volumes with full diffusional mixing \cite{Stange1998} and we shall not
further discuss it here. The most robust uniform oscillations, which we
consider, are characterized by the frequency $\omega \approx 2\pi / \tau$
and correspond to the single-group synchronization. As the result of a wave
bifurcation (also known as the Hopf bifurcation with a finite wave number 
\cite{Walgraef}), the first unstable modes are traveling waves with a
certain wavenumber $q_{0}$. Figure 3 shows the bifurcation diagram
in the parameter plane ($\tau_{p}$, $\beta$).
Note the presence of a codimension-2 bifurcation point
where the boundaries of the Hopf and the wave bifurcations join.

\begin{figure}
\includegraphics[width=86mm]{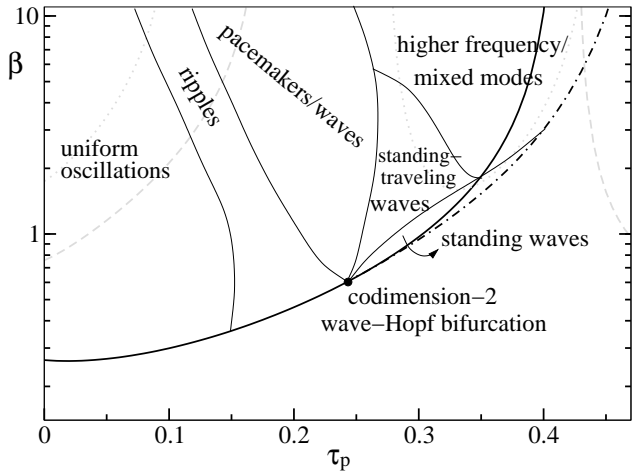}
\caption{Phase diagram ($\alpha_{0}=1$, $\alpha_{1}=1000$, $\kappa=10$,
$\gamma=10$, $c=100$, $D=1000$).
The Hopf bifurcation (solid line) and the wave bifurcation (dash-dotted line)
boundaries are displayed. Gray lines
show instability of the stationary state with respect to development of
uniform oscillations with two (dashed) and three (dotted) groups in the
well-mixed case. Lines separating parameter domains with different kinds of
patterns are hand-drawn, based on numerical simulations.}
\label{pd}
\end{figure}

To investigate nonlinear dynamics of the system, numerical simulations of
equations (1) have been performed \cite{Thesis}.
The explicit Euler integration method has
been used; no-flux boundary conditions were applied. Results of 1D
simulations are summarized in Fig. 3 and examples of typical observed
patterns are shown in Fig. 4. Standing waves (Fig. 4a) develop when the
boundary of the wave bifurcation (dash-dotted curve) is crossed and uniform
oscillations are observed above the boundary of the Hopf bifurcation. Near
the codimension-2 point, more complex behavior was found. This included
rippled oscillations (Fig. 4b), self-organized pacemakers (Fig. 4c) and
modulated traveling waves (Fig. 4d).
The observed patterns are similar to those previously found in
reaction-diffusion systems with the wave bifurcation \cite{Zhabotinsky}.
In the right upper corner of the
diagram in Fig. 3, higher frequency oscillations with several synchronous
groups take place.

Two-dimensional simulations of reaction-diffusion equations (1) with time
delay have been performed for selected parameter values. In 2D simulations,
spontaneously developing concentric waves (target patterns) and spiral waves
have been observed; target patterns were however unstable and evolved into
pairs of rotating spiral waves (Fig. 2c and Video 3 \cite{epaps}).
Complex wave regimes, which can be qualitatively characterized as turbulence of
standing waves, have also been observed (Fig. 2d and Video 4 \cite{epaps}).

\begin{figure}
\includegraphics[width=86mm]{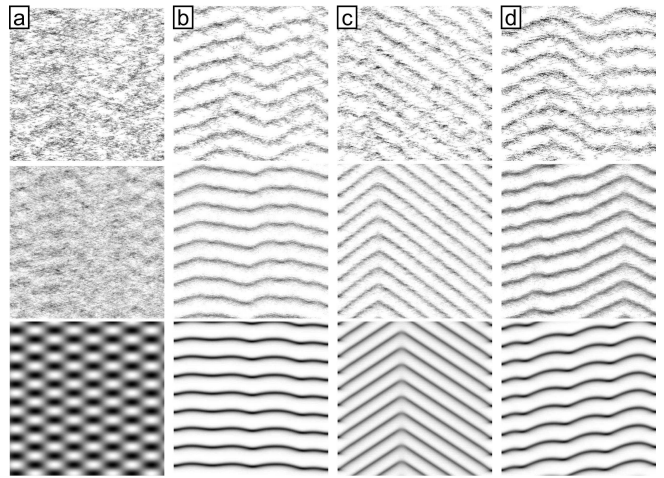}
\caption{Spatiotemporal patterns in a 1D system (in each panel,
the vertical axis is time, running down, and the
horizontal axis is the coordinate). The upper two rows are stochastic
simulations ($\sigma=0$) with concentrations $c=1$ and $c=10$, the
bottom row shows mean-field simulations with $c=100$.
(a) $\tau_{p}=0.3$, $\beta=95/c$,
(b) $\tau_{p}=0.14$, $\beta=260/c$,
(c) $\tau_{p}=0.22$, $\beta=600/c$,
and (d) $\tau_{p}=0.16$, $\beta=300/c$.
Other parameters as in Fig. 3; the system size shown is $L = 51$ $l_{diff}$.}
\label{patterns}
\end{figure}

The mean-field approximation is based on neglecting statistical fluctuations
in concentrations of reacting species \cite{Stange1998} and, therefore, it
should hold in the high concentration limit. In Fig. 4, two upper panel rows
display spatiotemporal patterns which are observed in stochastic
simulations with parameter values corresponding to the respective
mean-field simulations. To compare mean-field simulations with different
enzyme densities, the following property of equations (1) can be used:
introducing relative concentrations $\widetilde{n}_{0}=n_{0}/c$,
$\widetilde{n}_{1}=n_{1}/c$ and $\widetilde{m}=m/c$,
it can be noticed that they obey
the same equations, but with a rescaled coefficient
$\widetilde{\beta}=\beta c$.
Thus, essentially the same patterns are observed as long as
the parameter combination $\beta c$ remains constant. In the stochastic
simulations in Fig. 4, the coefficient $\beta$ has been increased to
compensate for a decrease in the enzyme concentration. For larger enzyme
concentrations, good agreement between mean-field predictions and
stochastic simulations has been found.
In the mean-field equations (1),
intramolecular fluctuations are not taken into account ($\sigma =0$ and
therefore each turnover cycle has the same fixed duration $\tau$).
Stochastic simulations have been, however, also performed when such
fluctuations were present. Synchronization
waves could still be found even at internal noise levels which corresponded
to the mean relative dispersion $\xi$ of turnover times of
about 10\% (with
$\xi = \left\langle \delta \tau^{2} \right\rangle^{1/2} / \tau \simeq \left(2\sigma \tau \right)^{1/2}$).

Although the emphasis in this Letter is on the phenomena in
two-dimensional enzymic arrays, analogous effects should be expected for
three-dimensional systems representing aqueous enzymic solutions. The linear
stability analysis, yielding Hopf and wave bifurcation boundaries (see
Fig. 3), is valid also for the 3D geometry. We have performed preliminary
stochastic simulations for thin solution layers with high enzyme
concentrations and could observe synchronization patterns
similar to those found for the enzymic arrays.

A product molecule, released by an enzyme, diffuses in the solution
until it either binds, as a regulatory molecule, to another enzyme or
undergoes a decay. Here, it should be taken into account that a regulatory
molecule can bind to an allosteric enzyme only at a certain binding site of
characteristic radius $R$.
Using the theory of diffusion-controlled reactions, the average
time $t_{transit}$ after which a regulatory product would find a binding
site of one of the enzymes can be roughly estimated \cite{Stange1998} as
$t_{transit} = 1/cDR$, if enzymes are uniformly distributed inside the reaction
volume with concentration $c$. Therefore, binding typically occurs within
the distance
$L_{corr}=\left( Dt_{transit} \right)^{1/2}=\left( cR \right)^{-1/2}$ from
the point where a molecule is released. Obviously, it can
only take place if the product molecule has not undergone decay until that
moment, i.e. if $\gamma t_{transit}<1$. This condition puts a restriction on
the enzyme concentration $c$, which must be higher than the critical
concentration $c^{\ast} = \gamma / DR$.
Choosing $\gamma =10^{3}$ s$^{-1}$, $D=10^{-5}$ cm$^{2}$s$^{-1}$ and
$R=10^{-7}$ cm, the critical enzyme
concentration is $c^{\ast} = 10^{15}$ cm$^{-3} = 10^{-6}$ M.
A similar estimate
can be obtained when enzymes are immobilized on a plane immersed into a
reactive solution; in this case the mean distance between the enzymes on
the plane should be less than
$l_{c}=\left( R l_{diff} \right)^{1/2}$ \footnote{Diffusion perpendicular
to the plane is considered as dilution
within a layer of effective thickness $\simeq l_{diff}$.}.
Although the required enzyme concentrations are relatively large, they are
within the range characteristic for biological cells (glycolytic
enzymes are present \cite{Hess1969} in a cell at even higher concentration
of more than 10$^{-5}$ M). The characteristic temporal period of
developing patterns is determined by the enzyme turnover time $\tau$, which
typically varies from milliseconds to seconds. The characteristic length
scale of developing wave patterns is determined by the diffusion length
$l_{diff}$, which can vary under these conditions from a fraction of a
micrometer to tens of micrometers.

Our analysis shows that spontaneous molecular synchronization of
allosteric product-activated enzymes can be observed in enzymic arrays.
Artificial arrays formed by immobilized protein machines (molecular motors)
are already used in experiments on active nanoscale transport
(see \cite{H-Hess}).
Many enzymes in biological cells are membrane-bound,
thus forming natural enzymic arrays.
Similar phenomena are possible in dense enzyme solutions.
In the study by Petty et al. \cite{Petty}, traveling
waves of NAD(P)H and proton concentrations with the wavelength of about a
micrometer were observed inside neutrophil cells.
These metabolic waves had the temporal period of about 300 ms,
which is by two orders of magnitude shorter than
the characteristic period of glycolytic oscillations in the cells
and lies closer to the time scales of turnover cycles of individual enzymes.
An intriguing question, requiring further detailed analysis,
is whether molecular synchronization waves may have already been seen in
these experiments.

Molecular synchronization waves are principally different from classical
concentration waves in reaction-diffusion systems. Under synchronization
conditions, internal conformational states of individual enzyme molecules
in their turnover cycles become strongly correlated. In optics, a similar
situation is found when a transition to coherent laser generation has taken
place. Our theoretical analysis may open a way to the investigations of a
new class of spatio-temporal pattern formation in chemically active
molecular systems.

The authors are grateful to M. Falcke and P. Stange for valuable
discussions. Financial support of Japan Society for the Promotion of
Science through a fellowship for research abroad (Y. T.) is
acknowledged.

\end{subequations}


\begin{thebibliography}{99}
\bibitem{Kinbara} K. Kinbara, T. Aida, \textit{Chem. Rev.} \textbf{105}, 1377 (2005).

\bibitem{Blumenfeld} L. A. Blumenfeld, A. N. Tikhonov, \textit{Biophysical Thermodynamics of Intracellular Processes: Molecular Machines of the Living Cell} (Springer, Berlin 1994).

\bibitem{Gerstein} M. Gerstein, A. M. Lesk, C. Chothia, \textit{Biochemistry} \textbf{33}, 6739 (1994).

\bibitem{Prost-review} F. J\"{u}licher, A. Ajdari, J. Prost, \textit{Rev. Mod. Phys.} \textbf{69}, 1269 (1997).

\bibitem{Lerch1} H.-Ph. Lerch, A. S. Mikhailov, B. Hess, \textit{Proc. Natl. Acad. Sci. (USA)} \textbf{99}, 15410 (2002).

\bibitem{Lerch2} H.-Ph. Lerch, R. Rigler, A. S. Mikhailov, \textit{Proc. Natl. Acad. Sci. (USA)} \textbf{102}, 10807 (2005).

\bibitem{Prost1} S. Ramaswamy, J. Toner, J. Prost, \textit{Phys. Rev. Lett.} \textbf{84}, 3494 (2000).

\bibitem{Prost2} P. Lenz, J.-F. Joanny, F. J\"{u}licher, J. Prost, \textit{Phys. Rev. Lett.} \textbf{91}, 108104 (2003).

\bibitem{Chen} H.-Y. Chen, \textit{Phys. Rev. Lett.} \textbf{92}, 168101 (2004).

\bibitem{Goldbeter} A. Goldbeter, \textit{Biochemical Oscillations and Cellular Rhythms} (Cambridge University Press, Cambridge 1996).

\bibitem{Stange1998} P. Stange, A. S. Mikhailov, B. Hess, \textit{J. Phys. Chem. B} \textbf{102}, 6273 (1998).

\bibitem{Stange1999} P. Stange, A. S. Mikhailov, B. Hess, \textit{J. Phys. Chem. B} \textbf{103}, 6111 (1999).

\bibitem{Ouyang} K. Sun, Q. Ouyang, \textit{Phys. Rev. E} \textbf{64}, 026111 (2001).

\bibitem{Gruler} M. Schienbein, H. Gruler, \textit{Phys. Rev. E} \textbf{56}, 7116 (1997).

\bibitem{epaps} See EPAPS Document No. E-PRLTAO-99-041730 for dynamical evolutions in the 2D simulations. For more information on EPAPS, see http://www.aip.org/pubservs/epaps.html .

\bibitem{Thesis} V. Casagrande, Doctoral thesis, Technical University, Berlin (2006),\\ http://opus.kobv.de/tuberlin/volltexte/2006/1273/ .

\bibitem{Walgraef} D. Walgraef, \textit{Spatio-Temporal Pattern Formation} (Springer, Berlin 1997).

\bibitem{Zhabotinsky} A. M. Zhabotinsky, M. Dolnik, I. R. Epstein, \textit{J. Chem. Phys.} \textbf{103}, 10306 (1995).

\bibitem{Hess1969} B. Hess, A. Boiteux, J. Kr\"{u}ger, \textit{Adv. Enzyme Regul.} \textbf{7}, 149 (1969).

\bibitem{H-Hess} H. Hess, G. D. Bachand, \textit{Materials Today} \textbf{8} (12, Suppl. 1), 22 (2005).

\bibitem{Petty} H. R. Petty, R. G. Worth, A. L. Kindzelskii, \textit{Phys. Rev. Lett.} \textbf{84}, 2754 (2000).

\end{thebibliography}
\end{document}